\documentclass{article}
\usepackage[cp1251]{inputenc}
\usepackage[T2A]{fontenc}
\usepackage{amsmath}
\usepackage{indentfirst}
\usepackage{amssymb}
\usepackage{literat}
\usepackage{graphicx}
\usepackage{cite}

\title{Influence of nonlinear dissipation and external perturbations onto transition scenarious to chaos in Lorenz-Haken system}  % Declares the document's title.
\author{A. V. Dvornichenko \\ Sumy State University, Rimskii-Korsakov Str., 2, 40000, Sumy, Ukraine}
\begin{document}
\maketitle
\section*{Abstract}

We study an influence of nonlinear dissipation and external perturbations onto
transition scenarious to chaos in Lorenz-Haken system. It will be show that
varying in external potential parameters values leads to parameters domain
formation of chaos realization. In the modified Lorenz-Haken system transitions
from regular to chaotic dynamics can be of Ruelle-Takens scenario, Feigenbaum
scenario, or through intermittency.

\section{Introduction}

One of the most actual tasks in the theory of nonlinear dynamical systems is a
setting of chaotic regime generation conditions and defining possibilities of
chaos control \cite{ieejournal,6fromiee}. It is well known, that in physical
applications for a transition from chaotic to periodic mode in multicomponent
systems different mechanisms are used. For example, in laser physics they are
negative feedback \cite{169}, angle between two crystals which are entered to
the Fabry-Perot cavity \cite{153}, full feedback intensity \cite{ieejournal},
intensity of activating \cite{156} (see \cite{bokaletti} and citations
therein). In this connection, from the theoretical point of view, an actual
task is to establish of chaos control and to determine transition characters
between chaotic and regular dynamics.

The main goal of this work is to study an influence of two additional
nonlinearityes that arise up in the chaotic system as a result of different
physical processes onto transitions character between regular and chaotic
dynamics. We shall consider a modified Lorenz-Haken model which
self-consistently can describes, for example, optical bistable systems
\cite{Haken}, systems of defects are in a solid
\cite{olemskoy_book,dikh_knyaz}, etc. Due to condition of commensurability for
all three modes relaxation times we shall set domains of system parameters of
chaos realization with a help of maximal Lyapunov exponents approach. We shall
obtain two different strange chaotic attractors and shall set possible
transitions to chaotic dynamics.

The paper is organized in the following manner. In Section 2 we present a model
of our system incorporating a nonlinear dissipation external perturbation
terms. Section 3 is devoted to the consideration of conditions for transition
to chaotic regime and to determination the main characteristics of strange
chaotic attractor. The main results and prospects for the future are presented
in the Conclusions.

\section{A model of a chaotic system}

A Lorenz-Haken model can be written in a form \cite{Haken}:
\begin{equation}
  \begin{split}
   &\dot{\eta}=-\eta/\tau_\eta+g_\eta h,\\
   &\dot{h}=-h/\tau_h+g_h\eta S,\\
   &\dot{S}=(r-S)/\tau_S-g_S\eta h.
  \end{split}
 \label{lor}
\end{equation}
Here a point means a derivative in time $\tau_\eta$, $\tau_h$, $\tau_S$ --
relaxation times of an order parameter $\eta(t)$, a conjugating field $h(t)$
and a control parameter $S(t)$, accordingly; $g_\eta$, $g_h$, $g_S$ -- positive
feed-back constants; $r$ -- pump intensity, measures the influence of
environment. First elements in the left hand of the systems (\ref{lor}) take
into account dissipation effects, peculiar to the synergetic systems.
Connection between the order parameter and conjugating field is linear (first
equation), in that time as an evolution of the conjugating field $h(t)$, and
control parameter $S(t)$ sets due to nonlinear feed-backs relations (second and
third equations, respectively). Principally that positive feed-backs which are
provided by constant $g_\eta$ and $g_h$ result in an increase in conjugating
field. These positive feed-backs are compensated by negative one due to
principle of Le-Shatel'e. As a result one has decreasing in control parameter
(see third equation in Eq.(\ref{lor})).

Let us start the analysis of the system (\ref{lor}) with passing to
dimensionless variables. Such transition is arrived due to measuring of time
$t$, order parameter $\eta$, conjugating field $h$, and control parameter $S$
in the followings units:
\begin{equation}
 \begin{split}
  &t\propto\tau_\eta \qquad \eta_e\propto\left(g_hg_S\right)^{-1/2},\\
  &h_e\propto\left(g_\eta^2g_hg_S\right)^{-1/2}, \qquad S_e\propto(g_\eta g_h)^{-1}.
 \end{split}
\nonumber
\end{equation}
In this time, dropping indexes, the system (\ref{lor}) becomes a form
\begin{equation}
  \begin{split}
   &\dot{\eta}=-\eta+h,\\
   &\sigma\dot{h}=-h+\eta S,\\
   &\varepsilon\dot{S}=(r-S)-\eta h,
  \end{split}
\label{lor2eqqq}
\end{equation}
where $\sigma\equiv\tau_h/\tau_\eta$, $\varepsilon\equiv\tau_S/\tau_\eta$. The
system (\ref{lor2eqqq}) is written in supposition of linear dependence for
order parameter relaxation time as $\tau_\eta(\eta)=const$. However, most real
physical systems are characterized by the nonlinear relaxation processes. In
this connection let us suppose that the order parameter relaxation time
$\tau_\eta$ increasing with increase in order parameter $\eta$ due to relation
\cite{JETP}:
\begin{equation}\label{f_k}
\tau_\eta(\eta)=1-\frac{\kappa}{1+\kappa+\eta^2},
\end{equation}
where $\kappa$ -- positive constant which play a role of an dissipation
intensity. From Eq.(\ref{f_k}) it is seen that relaxation time
$\tau_\eta(\eta)$ is independent of order parameter sign. Except for that,
relation (\ref{f_k}) has practical application namely, it designs the action of
optical filter, entered into the Fabry-Perot cavity of optically bistable
system (for example solid-state laser). Such acting provides establishment of
the stable periodic radiation (or time dissipative structure realization)
\cite{chaos}. Using a dependence (\ref{f_k}), first equation of
(\ref{lor2eqqq}) is generalized by an additional nonlinear term
$f_\kappa=-(\kappa\eta)/(1+\eta^2)$.

If one consider system in external field, then one need to take into
consideration external perturbations. In this article we shall model such
perturbations by the external potential $V_e$. Due to the standard catastrophe
theory such potential is given by three types of catastrophes \cite{Poston}. In
general case one has
\begin{equation}\label{V(E)}
V_e=A\eta+\frac{B}{2}\eta^2+\frac{C}{3}\eta^3+\frac{D}{4}\eta^4+\frac{E}{5}\eta^5,
\end{equation}
where $A$, $B$, $C$, $D$, $E$ -- parameters of the theory. For a catastrophe
$A_2$ one has $B=D=E=0$, for a catastrophe $A_3$: $C=E=0$ and for a catastrophe
$A_4$: $D=0$. The modified Lorenz-Haken system has the form
\begin{equation}
  \begin{split}
   &\dot{\eta}=-\eta+h+f_\kappa(\eta)+f_e(\eta),\\
   &\dot{h}=-h+\eta S,\\
   &\dot{S}=(r-S)-\eta h,
  \end{split}
\label{lor2eq}
\end{equation}
where we suppose $\sigma\simeq\varepsilon\simeq1$ and $f_e(\eta)\equiv-{\rm
d}V_e/{\rm d}\eta$. Variation in parameters of $f_\kappa(\eta)$ and $f_e(\eta)$
can to induce changing of the attractor topology in phase space.

\section{Chaos in a modified Lorenz-Haken system}

The system Eq.(\ref{lor2eq}) with nonlinear dependence of order parameter
relaxation time versus order parameter in a form (\ref{f_k}) ($\kappa\ne0$) but
with absence of additional perturbations ($V_e=0$) was considered in
\cite{chaos}. It was shown that in such a case the semirestricted domain of
system parameters (pump intensity $r$ and dissipation intensity $\kappa$) for
dissipative structure realization is formed. It was set that in a case of
linear dependence for order parameter relaxation time versus its values
($\kappa=0$) chaotic regime is not realize. In addition, it was found the chaos
domain, and it was determined conditions of chaos control. Finally it was
defined fractal and statistical properties of corresponding chaotic strange
attractor.

\begin{figure}[!t]
\centering
\includegraphics[width=100mm]{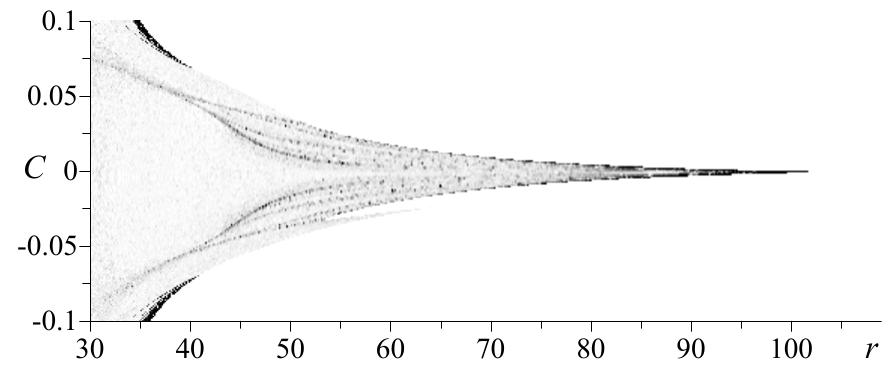}%
\caption{A Lypunov map of the modified Lorenz-Haken system (\ref{lor2eq}) at
$\kappa=25.0$ and $A=0.1$} \label{fig1}
\end{figure}

The main goal in this work is to study an influence of external perturbation,
on a regimes of transition to chaos in Lorenz-Haken system, generalized by
nonlinear relaxation time of order parameter in a form (\ref{f_k}). For
external perturbations we will use a potential of the fold catastrophe $A_2$,
i.e. $V_e\equiv A\eta+1/3C\eta^3$. To indicate a chaotic dynamic we shall use a
method of Lyapunov exponents which is provided by a Benettin algorithm
\cite{Benettin}. Due to this algorithm each of Lyaponov exponent (number is
defined by dimension of corresponding phase space) determines a speed of
convergence/divergence of any two initially nearby trajectories in a fixed
direction in corresponding phase space, starting from points $\vec{v}(t)$ and
$\vec{v}(t')$. The divergence/convergence of such trajectories is given by the
dependence $\delta\vec{v}(t)=\delta\vec{v}(t_0)e^{\Lambda_Mt}$, where
$\Lambda_M$ is a maximal (global) Lyapunov exponent, which is defined due to
relation \cite{Benettin}
 $$
 \Lambda_M\equiv
 \Lambda(\vec\delta(t_0))=
 \overline{\lim_{T\to\infty}}
 \frac{1}{T}
 \ln
 \left\|
 \frac{\delta{\vec{v}(t)}}{\delta\vec{v}(t_0)}
 \right\|.
 $$
Here one takes into account an upper limit and $\|\vec{v}\|$ is a norm;
$\vec{v}={\eta,h,S}$; $T$ -- full time. One can conclude that in a case
$\Lambda_i<0$, $i=1,2,3$, and accordingly $\Lambda_M<0$ all of phase
trajectories will coincide to fixed point (stable node or stable focus). At
$\Lambda_i<0$, $\Lambda_j<0$ and $\Lambda_M\equiv\Lambda_k=0$, $i\ne j\ne k$
$i,j,k=1,2,3$ phase trajectories will lie down on a stable limit circle
(dissipative structure). If $\Lambda_i<0$, $\Lambda_j=0$ and
$\Lambda_M\equiv\Lambda_k>0$, $i\ne j\ne k$ $i,j,k=1,2,3$ a dynamics of the
system is chaotic. A Lyapunov map of the modified Lorenz-Haken system
(\ref{lor2eq}) at $\kappa=25.0$ and $A=0.1$ is shown in Fig.\ref{fig1}.
Here by gradation of grey color the value of maximal (global) Lyapunov exponent
is shown versus pump intensity $r$ and parameter of an external potential $C$.
White color determines the domains of stable system behaviour (phase space is
characterized by a fixed point -- stable node or a stable focus). Grey color
marks the domains of time dissipative structure existence (more dark domains
correspond to the larger number of oscillation periods). Domains of chaos are
shown by a black colour. From Fig.\ref{fig1} it is seen, that in a case of
nonlinear term $C\eta^3$ absence in the external potential $V_e$ the existence
of the chaotic mode requires the large values of pump intensity. In case
$C\ne0:\ |C|\sim0.1$ the chaotic mode exists at $r\sim\kappa$.

\begin{figure}[!t]
\begin{center}
 \includegraphics[width=80mm]{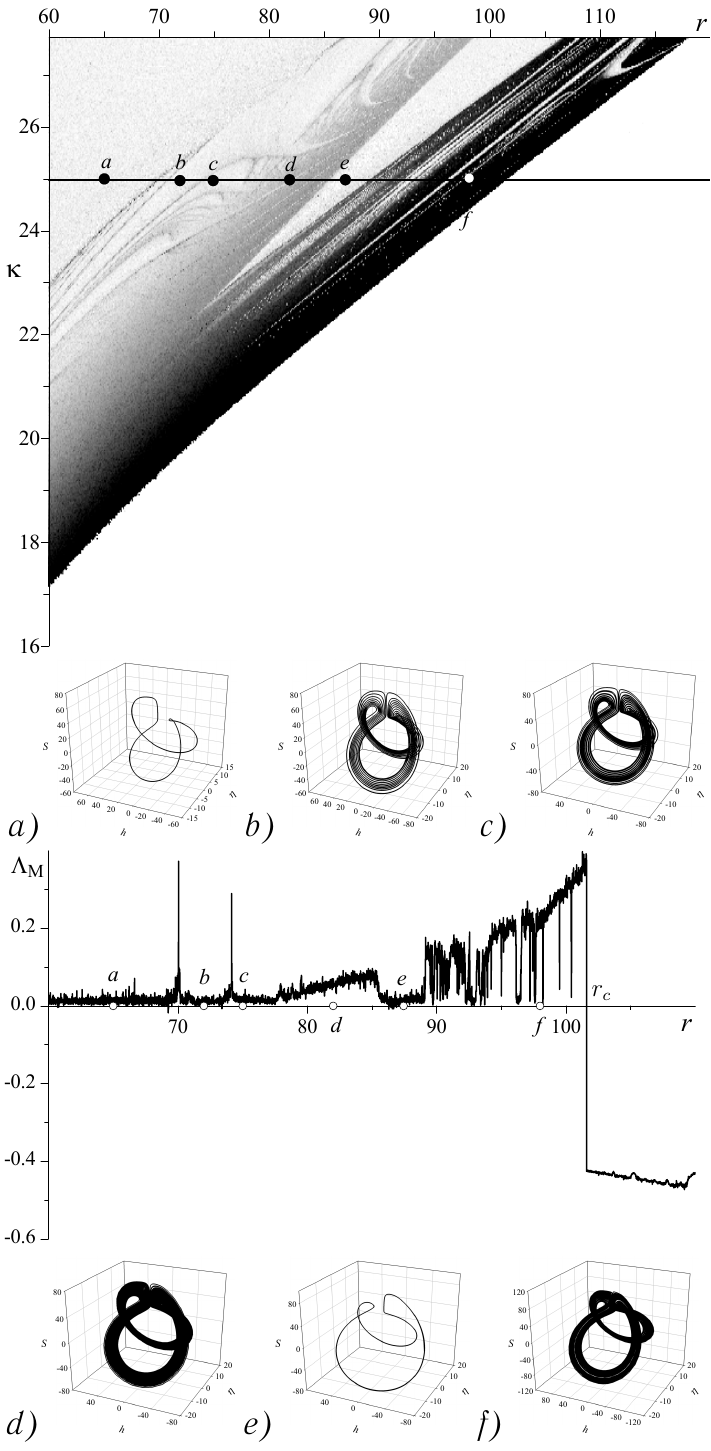}
\caption{Lyapunov map of the modified Lorenz-Haken system (\ref{lor2eq}) at
$A=0.1$ and $C=0.0$; dependence of maximal Lyapunov exponent at $A=0.1$,
$C=0.0$ and $\kappa=25.0$ and corresponding phase portraits}{\label{fig2}}
\end{center}
\end{figure}

Let us analyze a picture of dynamical regimes reconstruction for the system
(\ref{lor2eq}) with $C=0$ and $A=0.1$ in detail. Lyapunov map and corresponding
dependence of maximal Lyapunov exponent versus pump intensity at fixed value of
dissipation intensity $\kappa$ with characteristic phase portraits are shown in
Fig.{\ref{fig2}}.
\begin{figure}[!t]
\begin{center}
 \includegraphics[width=80mm]{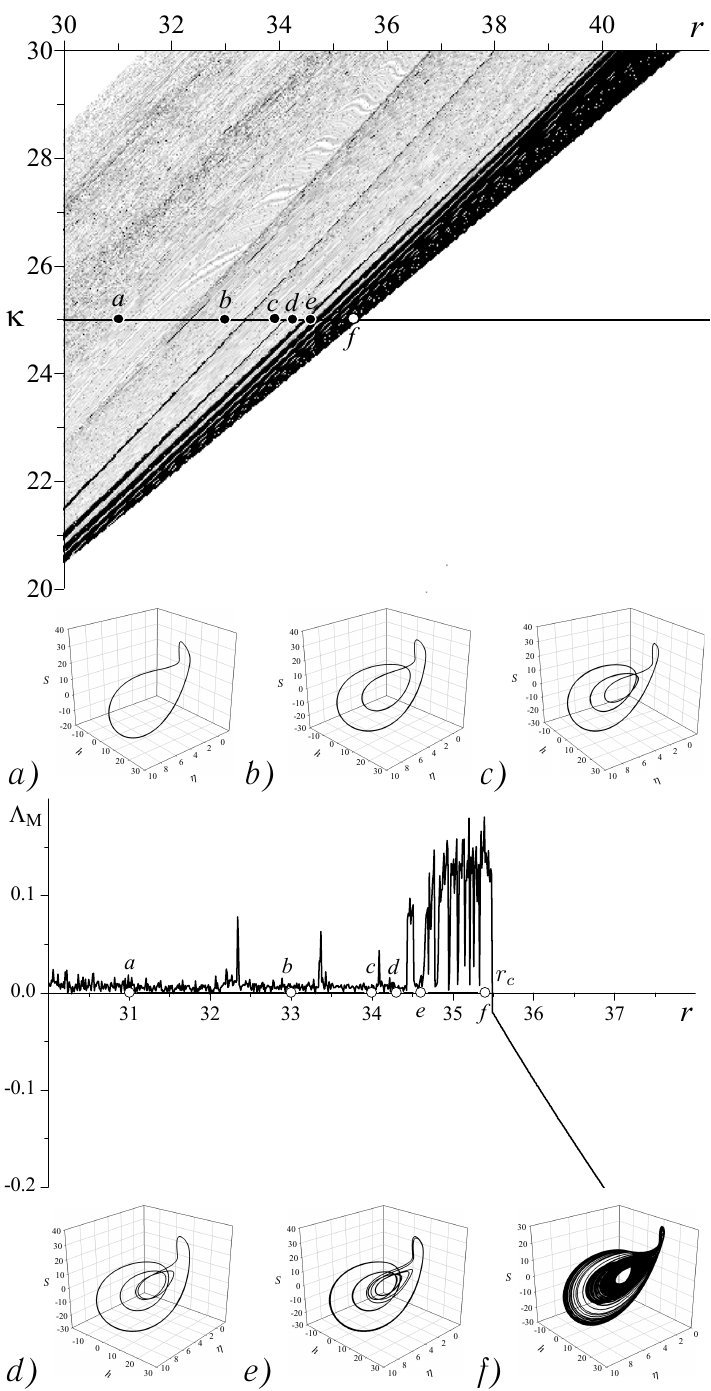}
\caption{Lyapunov map of the modified Lorenz-Haken system (\ref{lor2eq}) at
$A=0.1$ and $C=0.1$; dependence of maximal Lyapunov exponent at $A=0.1$,
$C=0.1$ and $\kappa=25.0$ and corresponding phase portraits}\label{fig3}
\end{center}
\end{figure}
Here due to earlier denoted scenario values of maximal Lyapunov exponent are
presented by gradation of grey color versus pump intensity and dissipation
intensity. It is necessary to note that dark curves (between \textit{a} and
\textit{b}, \textit{b} and \textit{c} in a Lyapunov map) in the domain of
dissipation structure existence (grey area) are determine the parameters values
of doubling period bifurcation. Below the map there is a dependence of maximal
Lyapunov exponent versus pump intensity $r$ at $A=0.1$, $C=0.0$ and
$\kappa=25.0$. A presence of two pronounced peaks in the domain of zero values
of maximal Lyapunov exponent (fluctuations around a zero are connected with an
error of numeral solution of systems (\ref{lor2eq})) determines the points of
doubling period bifurcation. Corresponding phase portraits illustrating
dissipative structure with one, $n$ and $m>n$ periods are shown with the help
of insertions \textit{a)}, \textit{b)} and \textit{c)}, accordingly. It is
principally, that as transition from \textit{a)} to \textit{b)}, as transition
from \textit{b)} to \textit{c)} characterizes by appearance of a few additional
harmonics. As it is seen from dependence $\Lambda_M(r)$ in a point \textit{d}
maximal Lyapunov exponent has positive value and phase space is characterized
by the irregular behaviour of trajectory. Increasing in pump intensity leads to
dissipative structure formation with one period (cf. phase portraits
\textit{d)} and \textit{e)}). Next increasing in $r$ leads to positive values
of $\Lambda_M$ and phase space is characterized by chaos existence
(corresponding phase portrait is shown with the help of insertion \textit{f)}.
So, one can conclude that at $A=0.1$, $C=0.0$ and $\kappa=25.0$ with an
increase in pump intensity $r$ a transition to chaotic regime occurs due to
Ruelle-Takens scenario, when only negligible number of doubling period
bifurcation leads to chaos \cite{shuster}. With a decreasing in $r$ (from
$r>r_c$ to $r<r_c$) maximal Lyapunov exponent takes positive value at $r=r_c$
in spontaneous manner. Corresponding transition to chaos takes a place through
intermittency \cite{Zaslavskiy}.

Next, let us consider the domain of chaos, shown in Fig.\ref{fig1} at $C\ne0$.
Lyapunov map at $A=0.1$ and $C=0.1$ is shown in Fig.\ref{fig3}. Below the map
as well as in previous case a dependence of maximal Lyapunov exponent versus
pump intensity at $A=0.1$, $C=0.1$ and $\kappa=25.0$ and corresponding phase
portraits are shown. Unlike to the previous case here with an increase in pump
intensity $r$ a successive complication of attractor due to doubling period
bifurcation is observed (see corresponding phase portraits in points
\textit{a}, \textit{b}, \textit{c}, \textit{d} and \textit{e}). Thus, in such a
case ($A=0.1$, $C=0.1$ and $\kappa=25.0$) an increasing in $r$ results to
transition to chaos due to Feigenbaum scenario \cite{shuster}. Chaotic
attractor is shown with the help of insertion \textit{f)}. As in a previous
case a decrease in $r$ leads to transition to chaos at $r=r_c$ through
intermittency \cite{Zaslavskiy}.

It is well known that dynamical systems can realize four types of attractors in
phase space, namely: non chaotic non strange attractor, chaotic non strange
attractor, strange non chaotic attractor and chaotic strange attractor. So, for
chaotic attractor one has ($\Lambda_M>0$) and for an strange one -- fractal
dimension of an attractor is fractional. In \cite{Sprott} it was shown that the
fractal dimension of an attractor which is realized in the dynamical system, is
determined with the help of Lyapunov exponents due to relation
 $$D\simeq1.5+0.5\sqrt{1-8\frac{\Lambda_{M}}{\Lambda_{min}}},$$
where $\Lambda_{min}$ is a minimal Lyapunov exponent. Thus, for considered
attractor in a point \textit{f} at $\kappa=25.0$, $r=98.0$, $A=0.1$ and $C=0$
(see Fig.\ref{fig2}) one has: $\Lambda_M=0.2774$, $\Lambda_{min}=-16.948$ and,
accordingly, $D\simeq2.032$. For the attractor at $\kappa=25.0$, $r=34.5$,
$A=0.1$ and $C=0.1$ (see Fig.\ref{fig3}) one has $\Lambda_M=0.16545$,
$\Lambda_{min}=-3.631$ and accordingly, $D\simeq2.084$. Thus, attractors in
Fig.\ref{fig2}f and Fig.\ref{fig3}f are strange and chaotic.

\section{Conclusions}

We have studied an influence of nonlinear dissipation and external
perturbations onto transition scenarious to chaos in Lorenz-Haken system.
Dissipation processes are defined due to nonlinear dependence of order
parameter relaxation time $\tau_\eta$ versus its values. External perturbations
are modeled by a potential of fold catastrophe $A_2$.

From a physical view point we have considered an absorptive optical bistability
system \cite{Haken}. At that time used nonlinear dissipation relation related
to the possibility of the additional medium in the Fabry-Perot cavity
(phthalocyanine fluid, gases $SF_6$, $BaCl_3$, and $CO_2$ \cite{Haken}) to
absorbing signals with weak intensities. Meanwhile, external perturbations
model an influence of optical modulator which sets additional interphotons
interaction processes in the Fabry-Perot cavity.

We have considered a case of commensurability of relaxation times for order
parameter, conjugated field and control parameter. It has been shown that
varying in external potential parameters values leads to parameters domain
formation with $r\sim\kappa$ and with $r$ by order of magnitude greater than
$\kappa$ of chaos realization. In considered system transitions from regular to
chaotic dynamics can be of Ruelle-Takens scenario, Feigenbaum scenario, or
through intermittency.

\end{document}